\documentclass{ws-rv961x669}
\usepackage[square]{ws-rv-van} 
\usepackage{ws-rv-thm} 
\usepackage{subfigure}
\newcommand{\beq}{\begin{equation}}
\newcommand{\eeq}{\end{equation}}
\newcommand{\bdm}{\begin{displaymath}}
\newcommand{\edm}{\end{displaymath}}
\newcommand{\ba}{\begin{array}}
\newcommand{\ea}{\end{array}}
\newcommand{\bn}{\begin{eqnarray}}
\newcommand{\en}{\end{eqnarray}}
\newcommand{\bns}{\begin{eqnarray*}}
\newcommand{\ens}{\end{eqnarray*}}
\newcommand{\bc}{\begin{center}}
\newcommand{\ec}{\end{center}}
\newcommand{\bit}{\begin{itemize}}
\newcommand{\eit}{\end{itemize}}
\newcommand{\btab}{\begin{tabular}}
\newcommand{\etab}{\end{tabular}}

\newcommand{\bls}[1]{\boldsymbol{#1}}
\makeindex
\begin{document}
\chapter[Octupole Hamiltonian]{The octupole collective Hamiltonian.\\ 
Does it follow the example of the quadrupole case?}
\author{Stanis{\l}aw G. Rohozi\'nski\footnote{Dedicated to the Memory of the late Professor Walter Greiner.
}}
\address{Institute of Theoretical Physics, Faculty of Physics, University of Warsaw, Pasteura 5, 02-093 Warsaw, Poland \\
Stanislaw-G.Rohozinski@fuw.edu.pl}
\author[S.G.Rohozi\'nski and L. Pr\'ochniak]{Leszek  Pr\'ochniak}
\address{Heavy Ion Laboratory, University of Warsaw,\\ Pasteura 5A, 02-093 Warsaw, Poland \\ prochniak@slcj.uw.edu.pl}
\begin{abstract}
A general form of the octupole collective Hamiltonian is introduced and analyzed based on fundamental tensors in the seven-dimensional tensor space.
Possible definitions of intrinsic frames of reference possessing cubic symmetry for the octupole tensor are considered. Cubic intrinsic octupole coordinates or
deformations are introduced. Shapes of the octupoloid are investigated. The octupole collective Hamiltonian is expressed in intrinsic coordinates.
An intrinsic angular momentum carried by the octupole vibrations is discovered. Small oscillations about an axially-symmetric pear shape are analyzed.
Formulation of a unified quadrupole-octupole collective model is discussed.  
\end{abstract}
\body
\section{Introduction}\label{IN}
The idea of attributing a definite multipolarity to nuclear collective excitations came from the nuclear liquid-drop model.
It was known before the discovery of atomic nuclei that the normal modes of the surface vibrations of a spherical drop of incompressible liquid have definite 
multipolarities \cite{Ray879}. Many years later, when atomic nuclei were already known, Siegfried Fl\"ugge \cite{Flu41} connected the eigenfrequencies
of the surface vibrations of a nuclear liquid drop with the excitation energies of low-lying states of even-even nuclei. These were the beginnings of the collective model.

The most important collective mode in nuclear structure physics is that with multipolarity $\lambda = 2 $, because it concerns the lowest excited states in even-even nuclei. These states are well known experimentally and theoretical methods describing them are well developed. One such method, applicable to even-even nuclei only, is a description by means of the Schr\"odinger equation in a collective space. Initiated by Aage Bohr \cite{Boh52} a long time ago, it is very effective and is still in use. The Bohr Hamiltonian became the label of the model. When applying the method one usually starts with a general classical collective Hamiltonian \cite{Bel65,Kum67}.
Its form has been extracted from various microscopic many-body models through methods of the ``Adiabatic Time-Dependent Hartree-Fock-Bogolyubov" (ATDHFB) type (see e.g. Chapt.~12 in Ref.\cite{Rin80}) and then quantized by means of the Podolsky-Pauli prescription \cite{Pod28,Pau33,Hof72}. Such a quasi-classical approach is still used even today. It would be good to use purely quantum methods to describe the collective states. Therefore, the Generator Coordinate Method (GCM) has been recently employed \cite{Ben08,Rod10}. It leads to integral equations instead of differential ones (see e.g. Chapt.~10 in Ref.\cite{Rin80}). However, in an approximation to the GCM called the Gaussian Overlap Approximation (GOA) the Hill-Wheeler integral equations for collective motion can be reduced back to differential equations with a collective Hamiltonian \cite{Roh12}. 

Apart from the positive-parity quadrupole states, negative-parity levels have been observed for a long time \cite{Asa53} in the low-energy spectra of even-even nuclei. Multipolarity $\lambda =3$
has been attributed to such states. Information on the octupole states has been collected over many years, see Refs.~\cite{Roh88} and \cite{But96} for reviews. Recently, measurements of a static octupole deformation in radium ($^{224}$Ra) and barium ($^{144}$Ba) isotopes have been reported \cite{Gaf13,Buc16}. However, data on the octupole states are not so rich as in the case of quadrupole excitations. This is perhaps connected with experimental and interpretational difficulties. 
A common opinion seems to be that the quadrupole collective Hamiltonian stands for a pattern for higher multipolarities. However, it is not so easy. The theory of the octupole degrees of freedom is much more complicated than that of the quadrupole ones, or rather the quadrupole case is exceptionally simple. This will be demonstrated further in subsequent Sections. In \sref{GCH} a general form of
the octupole collective Hamiltonian is introduced and discussed in comparison with that of the quadrupole one. Possible definitions of intrinsic frames of reference for the octupole tensor in  analogy to that for the quadrupole tensor are considered in \sref{IF}. The octupole deformation parameters are introduced and illustrated. The octupole Hamiltonian is expressed in intrinsic 
coordinates in \sref{IHAM}. Its approximate form for small oscillations around a pear shape is given in \sref{AS}. In Conclusion, in \sref{CONCL}, a draft of a unified quadrupole-octupole collective model is recapitulated.

\section{General form of collective Hamiltonians}\label{GCH}
The idea of collective models consists generally in describing some complex (collective) states of a many-body system through the substitution of the coordinates of many particles by a relatively small number
of collective variables. An appropriate choice of these collective variables decides the success of the model. Constructing the collective models in question here for the description of low-lying collective states of even-even nuclei one takes the spherical tensors $\bls{\alpha }_\lambda $ as collective variables. These tensors are transformed according to irreducible $(2\lambda +1)$-dimensional representations D$^\lambda $ of the O(3) group of orthogonal transformations in the physical space. The spherical components $\alpha _{\lambda \mu }$ 
($\mu = -\lambda ,\dots ,\lambda $) of $\bls{\alpha }_\lambda $ in the laboratory frame, U$_\mathrm{lab}$, are the collective laboratory coordinates. It is assumed that the tensor 
$\bls{\alpha }_\lambda $ is electric i.e. has parity $(-1)^\lambda $  and real, which means that its components fulfill the relation $\alpha^*_{\lambda \mu }= (-1)^\mu \alpha _{\lambda  -\mu }$. The differential operators
\footnote{Units $\hbar =c = 1$ are used here.}
$-i\partial/\partial\alpha ^*_{\lambda \mu }$ play the role of the momenta canonically conjugate to the coordinates $\alpha _{\lambda \mu }$. The angular momentum operators or the O(3) 
generators in the collective space fulfill characteristic commutation relations with the coordinates and momenta (see e.g. Appendix {\it A.1.} in Ref.~\cite{Pro09}). The following vector operator:
\beq\label{angmom}
L^{(\lambda )}_{1\mu }(\bls{\alpha }_\lambda )= (-1)^\lambda \sqrt{\frac{\lambda (\lambda +1)(2\lambda +1)}{3}}\sum_{\kappa \nu }(\lambda \kappa \lambda \nu |1\mu )
\alpha _{\lambda\kappa } \frac{\partial}{\partial\alpha ^*_{\lambda \nu }}
\eeq
fulfills such commutation relations and thus plays the role of the collective angular momentum \cite{Pro09,Roh82}.

The nuclear collective system is defined by a collective Hamiltonian $H_\lambda (\bls{\alpha }_\lambda )$.
It is assumed that $H_\lambda (\bls{\alpha }_\lambda )$ has the following properties:
\begin{arabiclist}
\item It is a second-order differential operator (like the Schr\"odinger operator),
\item it is real, i.e. $H_\lambda (\bls{\alpha }_\lambda )= H^*_\lambda (\bls{\alpha }_\lambda )$ (it describes even-even nuclei only),
\item it is a scalar with respect to the orthogonal group O(3) (rotations and inversion in the physical space) or commutes with the angular momentum operators of \eref{angmom},
\item it is Hermitian with a scalar weight $W(\bls{\alpha }_\lambda ) \ge 0$,
\item it possesses the lowest eigenvalue (positive kinetic energy),
\item it is an isotropic function of the coordinates, i.e. it does not depend on any other tensor quantities.
\end{arabiclist}
No other assumptions are needed at this stage (cf. Ref.~\cite{Nis14}).
The most general form of $H_\lambda (\bls{\alpha }_\lambda )$ with properties (1) -- (6) given above reads
 \cite{Pro09,Roh12}:
\beq\label{haml}
 H_\lambda (\bls{\alpha }_\lambda )= -\frac{1}{2 W(\bls{\alpha }_\lambda  ) }\sum_{\mu ,\nu}\frac{\partial}{\partial\alpha _{\lambda \mu}}
W(\bls{\alpha }_\lambda ) B^{-1}_{\lambda \mu \lambda \nu}(\bls{\alpha }_\lambda  )\frac{\partial}{\partial\alpha _{\lambda \nu}} + V(\bls{\alpha }_\lambda  ) ,
\eeq
The Hamiltonian is determined by three quantities: two real scalar functions --- weight $W(\bls{\alpha }_\lambda )$ and potential  $V(\bls{\alpha }_\lambda  )$, 
and a symmetric $(2\lambda +1)\times (2\lambda +1)$ matrix of real so-called inverse inertial functions 
$B^{-1}_{\lambda \mu \lambda \nu}(\bls{\alpha }_\lambda  )$. Since the matrix is transformed under the O(3) group like a product of two tensors $\bls{\alpha }_\lambda $ it is called a symmetric bitensor. All these three quantities depend on no other tensors but $\bls{\alpha }_\lambda $  and therefore they are isotropic tensor fields in the $(2\lambda +1)$-dimensional collective space. They can either be calculated from microscopic many-body models
or fitted to experimental data. When the Hamiltonian is obtained by quantization of its classical counterpart the weight is 
$W(\bls{\alpha }_\lambda ) =\sqrt{\det{B_{\lambda \mu \lambda \nu }(\bls{\alpha }_\lambda )}} $. 

In order to investigate possible structures of the inverse inertial bitensor it is convenient to express it as a set of (single) tensors $\bls{T}_{2l }(\bls{\alpha }_\lambda )$
($l = 0,\dots ,\lambda $), namely
\beq\label{bll}
B^{-1}_{\lambda \mu \lambda \nu}(\bls{\alpha }_\lambda  ) = \sum_{l  =0}^{\lambda }(\lambda \mu \lambda \nu |2l\  m )T_{2l\  m }(\bls{\alpha }_\lambda )
\eeq
Assumption (6), that the $\bls{T}_{2l }(\bls{\alpha }_\lambda )$ are isotropic functions of $\bls{\alpha }_\lambda $ is essential here. Then, an arbitrary tensor 
$\bls{T}_\Lambda (\bls{\alpha }_\lambda )$ can be expressed in the following form:
\beq\label{tstau}
\bls{T}_\Lambda (\bls{\alpha }_\lambda )= \sum_{k=1}^{k_{\lambda \Lambda }}\sigma ^{(\Lambda )}_k (\bls{ \alpha }_\lambda )\bls{\tau }_{k\Lambda }(\bls{\alpha }_\lambda )
\eeq
by a number of definite fundamental tensors $\bls{\tau }_{k\Lambda } (\bls{\alpha }_\lambda ) $ 
for given $\lambda $ and $\Lambda $,
and arbitrary scalar coefficients $\sigma ^{(\Lambda )}_k (\bls{\alpha }_\lambda )$ (see Appendix A in Ref.~\cite{Pro09}). The original Bohr inverse inertial bitensor \cite{Boh52} 
has the following form:
\beq\label{bibt}
B^{-1}_{\lambda \mu \lambda \nu}(\bls{\alpha }_\lambda  ) = (\lambda \mu \lambda \nu |00 ) (-1)^\lambda \sqrt{2\lambda +1}\frac{1}{B_\lambda } = \frac{(-1)^\mu \delta _{\mu -\nu }}{B_\lambda }
\eeq
where $B_\lambda $ is a constant mass parameter.

In the case of a general quadrupole $(\lambda =2)$ collective Hamiltonian it is well known that the inverse inertial bitensor can have at most six independent components
out of fifteen possible ones \cite{Kum67,Pro09}. It is interesting to know whether there are similar restrictions for the components of the inverse inertial bitensor in the case of an octupole 
$(\lambda =3 )$ collective Hamiltonian. Unfortunately, the things are much more complicated in that case. The symmetric bitensor $B^{-1}_{3 \mu 3 \nu}(\bls{\alpha }_3  )$ has twenty-eight components. 
Can they all be independent? To answer this question one should analyze Eqs. (\ref{bll}) and (\ref{tstau}) for $\lambda =3$.

According to \eref{bll} the even (positive-parity) symmetric octupole bitensor, $B^{-1}_{3 \mu 3 \nu}(\bls{\alpha }_3  )$, can be replaced with four tensors $\bls{T}_{2l}(\bls{\alpha }_3 )$ ( $l = 0,\dots ,3$).
These tensors should be even isotropic functions of $\bls{\alpha }_3$. The tensor fields in the space of the octupole coordinates are built out of some twenty-six elementary tensors 
\beq\label{elten}
\bls{t}^{(n)}_l  =[\underbrace{\bls{\alpha}_3\times\ \dots\ \times\bls{\alpha}_3}_{n}]_l 
\eeq 
(square brackets $[\dots ]_l$ stand for vector coupling to rank $l$)
for $l < 3n $ and related to each other by about two hundred syzygies or relations in the form of rational integral functions \cite{Roh78}.
All the elementary tensors split into two groups with positive and negative spin-parity $(-1)^{l+n}$, respectively.  Independent fundamental tensors $\bls{\tau }_{k\Lambda }(\bls{\alpha }_3 )$ from 
\eref{tstau} are constructed by alignment of the elementary tensors of \eref{elten}.  The positive spin-parity elementary and even (positive parity) fundamental tensors needed to construct the inverse inertial bitensor in question are listed in \sref{ET}. There are twenty-eight relevant fundamental tensors altogether and thus, all twenty-eight components of the bitensor 
$B^{-1}_{3\mu 3\nu }(\bls{\alpha }_3 )$ can be arbitrary tensor fields
in the octupole collective space. No additional relations between the components need appear. Twenty-eight scalars  
$\sigma ^{(\Lambda )}_k (\bls{\alpha }_3 )$ for $k = 1,\dots ,k_{3\Lambda }$ and
$\Lambda = 0,\ 2,\ 4,\ 6$ are functions of the four elementary scalars, \eref{eet}.

\section{Intrinsic frame and intrinsic coordinates}\label{IFIC}
\subsection{Intrinsic frames of reference}\label{IF}

Obviously, the tensor $\bls{\alpha }_\lambda $ can be represented by different sets of coordinates in different frames of reference. 
As already stated in \sref{GCH}, the $\alpha _{\lambda \mu}$ are the components of $\bls{\alpha }_\lambda $ in  the frame U$_\mathrm{lab}$.
Let us take another frame, say U$_\mathrm{in}$, the orientation of which with respect to the laboratory frame U$_\mathrm{lab}$ is given by the Euler angles $\omega =(\varphi ,\vartheta ,\psi )$.
We will not use the spherical components of $\bls{\alpha }_\lambda $ in  the frame U$_\mathrm{in}$. Instead, we shall use their real and imaginary parts, $a_{\lambda k}$ and $b_{\lambda k}$,
respectively, defined in the standard way (cf e.g. Eq.~A.7 in Ref.~\cite{Pro09}). 
Then, the transformation
rule between the corresponding components of the tensor $\bls{\alpha }_\lambda $ with respect to  U$_\mathrm{lab}$ and U$_\mathrm{in}$, respectively, takes the following form (cf. Ref.~\cite{Roh82}):
\beq\label{retr}
\alpha _{\lambda \mu}=  D^{\lambda (+)}_{\mu 0}(\omega )a_{\lambda 0} + \sum_{k =1,2,3}[ D^{\lambda (+)}_{\mu k}(\omega )a_{\lambda  k}+D^{\lambda (-)}_{\mu k}(\omega )b_{\lambda  k}]
\eeq
where 
\bn\label{semicart}
&& D^{\lambda (+)}_{\mu k}(\omega )=\frac{1}{\sqrt{2(1+\delta_{k 0}{)}}}[ \mathcal{D}^\lambda _{\mu k}(\omega) +(-1)^k\mathcal{D}^\lambda _{\mu -k}(\omega)]\nonumber \\
&& D^{\lambda (-)}_{\mu k}(\omega )=\frac{i}{\sqrt{2}}[ \mathcal{D}^\lambda _{\mu k}(\omega) -(-1)^k\mathcal{D}^\lambda _{\mu -k}(\omega)]
\en
are the semi-Cartesian Wigner functions.
The Bohr-Mottelson definition of the Wigner functions, $\mathcal{D}^\lambda _{\mu k}(\omega)$, is used \cite{Boh69}.

For the frame U$_\mathrm{in}$ to be the intrinsic frame, three properly chosen conditions for the coordinates $a_{\lambda k}$ and $b_{\lambda k}$  should be given, namely
\beq\label{intr}
\Omega _i(a_{\lambda k}(\omega ,\alpha _{\lambda \mu}),b_{\lambda k}(\omega ,\alpha _{ \lambda \mu}))= \Omega  _i(\omega ,\alpha _{\lambda \mu})=0
\eeq
for $i=1,\ 2,\ 3$, which determine the three Euler angles,
$\omega =(\varphi ,\vartheta ,\psi )$ as functions of the laboratory coordinates and, in this way, impart the status of intrinsic coordinates to them. The remaining $2\lambda -2$ independent coordinates are usually called deformations.

The concept of an intrinsic (body-fixed) frame of reference is connected with the descriptions of collective states from the very beginning \cite{Boh52}.
The principal axes of the tensor $\bls{\alpha }_2$ are taken as the intrinsic axes in the quadrupole, $\lambda =2$, case. According to \eref{cc2} from  Appendix \ref{CC} the well-known
definition of the intrinsic frame is $g_{2s} =0$ for $s =x,y,z$. It is seen from Appendix \ref{SCF} that the frame U$_\mathrm{in}$ is then O$_\mathrm{h}$-symmetric. 
The two remaining intrinsic coordinates, $a_{20} = \beta \cos{\gamma }$ and $a_{22}=\beta \sin{\gamma }$, are usually parametrized by the well-known deformation parameters $\beta $ and $\gamma $.

The problem of the intrinsic frame for the octupole tensor, $\bls{\alpha }_3$, has appeared to be less transparent than that for $\lambda =2$. The tensor has no principal axes and thus no
obvious intrinsic frame. Early attempts to define an intrinsic frame failed ({\it cf} Ref.~\cite{Dav65}). This was, perhaps, the reason why the  intrinsic frame of reference was for a long time determined 
with the quadrupole tensor and octupole coordinates were treated as intrinsic coordinates with respect to that frame (cf. Ref.~ \cite{Roh82}). And what shall we do when any quadrupole tensor is not at our disposal?

In order to define an intrinsic frame for $\lambda =3$, having the natural symmetry O$_\mathrm{h}$, it is convenient to decompose the tensor representation D$^3$ of the O(3) group onto  
the O$_\mathrm{h}$ irreps A$^-_2$, F$^-_1$ and F$^-_2$, respectively (see Appendix~\ref{OH}). The corresponding representations will be called the octupole cubic Wigner functions $A_{\mu}(\omega )$, $F_{\mu s}(\omega ),\ (s=x,y,z)$ and $G_{\mu s}(\omega ),\ (s=x,y,z)$, respectively. They are the following combinations of the semi-Cartesian Wigner functions:
\bn\label{cubic}
&& A_{\mu}(\omega )=D^{3(-)}_{\mu 2}(\omega ) \nonumber \\
&& F_{\mu x}(\omega )=\sqrt{\frac{3}{8}}D^{3(+)}_{\mu 1}(\omega )-\sqrt{\frac{5}{8}}D^{3(+)}_{\mu 3}(\omega ), \nonumber \\
&& F_{\mu y}(\omega )=\sqrt{\frac{3}{8}}D^{3(-)}_{\mu 1}(\omega )+\sqrt{\frac{5}{8}}D^{3(-)}_{\mu 3}(\omega ) , \nonumber \\
&& F_{\mu z}(\omega )= D^{3(+)}_{\mu 0}(\omega ) \nonumber \\
&& G_{\mu x}(\omega )=\sqrt{\frac{5}{8}}D^{3(+)}_{\mu 1}(\omega )+\sqrt{\frac{3}{8}}D^{3(+)}_{\mu 3}(\omega ) \nonumber \\
&& G_{\mu y}(\omega )= -\sqrt{\frac{5}{8}}D^{3(-)}_{\mu 1}(\omega )+\sqrt{\frac{3}{8}}D^{3(-)}_{\mu 3}(\omega ) \nonumber \\
&& G_{\mu z}(\omega )= D^{3(+)}_{\mu 2}(\omega ).
\en
The cubic Wigner functions form a unitary set. 
When using them the transformation rule of \eref{retr} takes the following form:
\beq\label{labcub}
\alpha _{3\mu}(\omega ,b ,f )= A_{\mu}(\omega )b + \sum_{s=x,y,z}[F_{\mu s}(\omega)f_{s} + G_{\mu s}(\omega)g_{s}]
\eeq
where $b$, $f_s$ and $g_s$ are the octupole cubic coordinates in the frame U$_\mathrm{in}$, proposed in Ref.~\cite{Roh90} and defined in Appendix~\ref{CC}. 
 
It is seen that we have two possible definitions of the O$_\mathrm{h}$-symmetric frame of reference. We can take \eref{intr}  in the two following alternative forms:
either $g_s =0$ or $f_s =0$ for $s =x,y,z$. Here, we shall explore the former definition. Both of them are briefly discussed in Ref.~\cite{Roh17}.
We see from \eref{labcub} that in the former case the following relation between the spherical laboratory coordinates and the Euler angles and the octupole deformations
holds:
\beq\label{intrtolabf}
\alpha _{3\mu}(\omega ,b ,f )= A_{\mu}(\omega )b + \sum_{s=x,y,z}F_{\mu s}(\omega)f_{s}
\eeq
The Jacobian of the transformation $\alpha _\mu \to \omega ,b, f_s$ is equal to
\bn\label{jacf}
 D_f(\vartheta ,b , f_{x},f_{y},f_{z})
& = & 8\sin{\vartheta}\left[b\left(b^2-\frac{15}{16}\left(f_{x}^2+f_{y}^2+f_{z}^2\right)\right) +\frac{15}{8}\sqrt{\frac{15}{16}}f_{x}f_{y}f_{z}\right] \nonumber \\
& = & 8\sin{\vartheta}d_f (b, f_s)
\en
The transformation (\ref{intrtolabf}) is reversible for the deformations contained aside from the hyper-surface $D_f(\vartheta ,b , f_{x},f_{y},f_{z})=0$, where ambiguities appear arising from the symmetries of the shape. The F$^-_1$-covariant (vector) deformations ($f_{x},f_{y},f_{z}$)
are transformed under the O$_{\mathrm{h}}$ transformations of the intrinsic frame like the Cartesian coordinates $x,y,z$ of a position vector.
\begin{figure}
\centerline{\includegraphics[width=4.5cm]{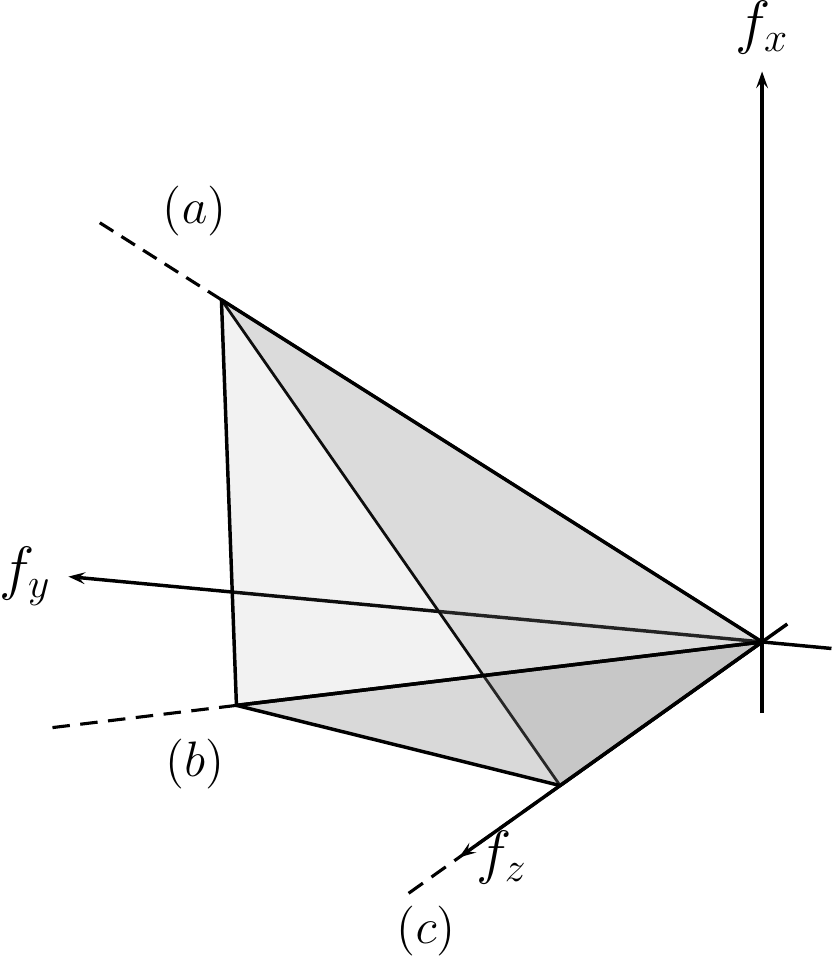}}
\caption{Infinite triangular pyramid mapping out the three-dimensional region of octupole vector deformations ($f_x, f_y, f_z$) with the vertex at point (0,0,0). The three pyramid edges are: 
(a) principal diagonal of octant $f_x,f_y,f_z\geq 0$, (b) diagonal of plane $f_yf_z$, (c) axis $f_z$.}
\label{fig_pyr}
\end{figure}
It follows directly from the O$_{\mathrm{h}}$ symmetry that it is sufficient to consider, for instance, the 
region  $0\leq f_{x}\leq f_{y}\leq f_{z}$ of vector deformations forming an infinite triangular pyramid in three-dimensional space (see \fref{fig_pyr}). The forty-eight pyramids obtained by all the O$_{\mathrm{h}}$ transformations fill up the entire space.  
Deformation $b$ supplements the space of vector deformations to the four-dimensional space. It is invariant under rotations $R_1$ and $R_3$ and changes sign under $R_2$ and inversion $P$ together
with the corresponding transformations of the vector deformations. Therefore, there are no additional restrictions on values of $b$ (see \fref{fig_pmbfz} below). The case when 
$f_x^2 +f_y^2 +f_z^2 = 0$ is an exception: it is sufficient then to consider values $b\geq 0$.  

\begin{figure}[ht]
\centerline{
  \subfigure[$b=0.5, f_x =f_y = f_z =0$]
  {\includegraphics[width=2in]{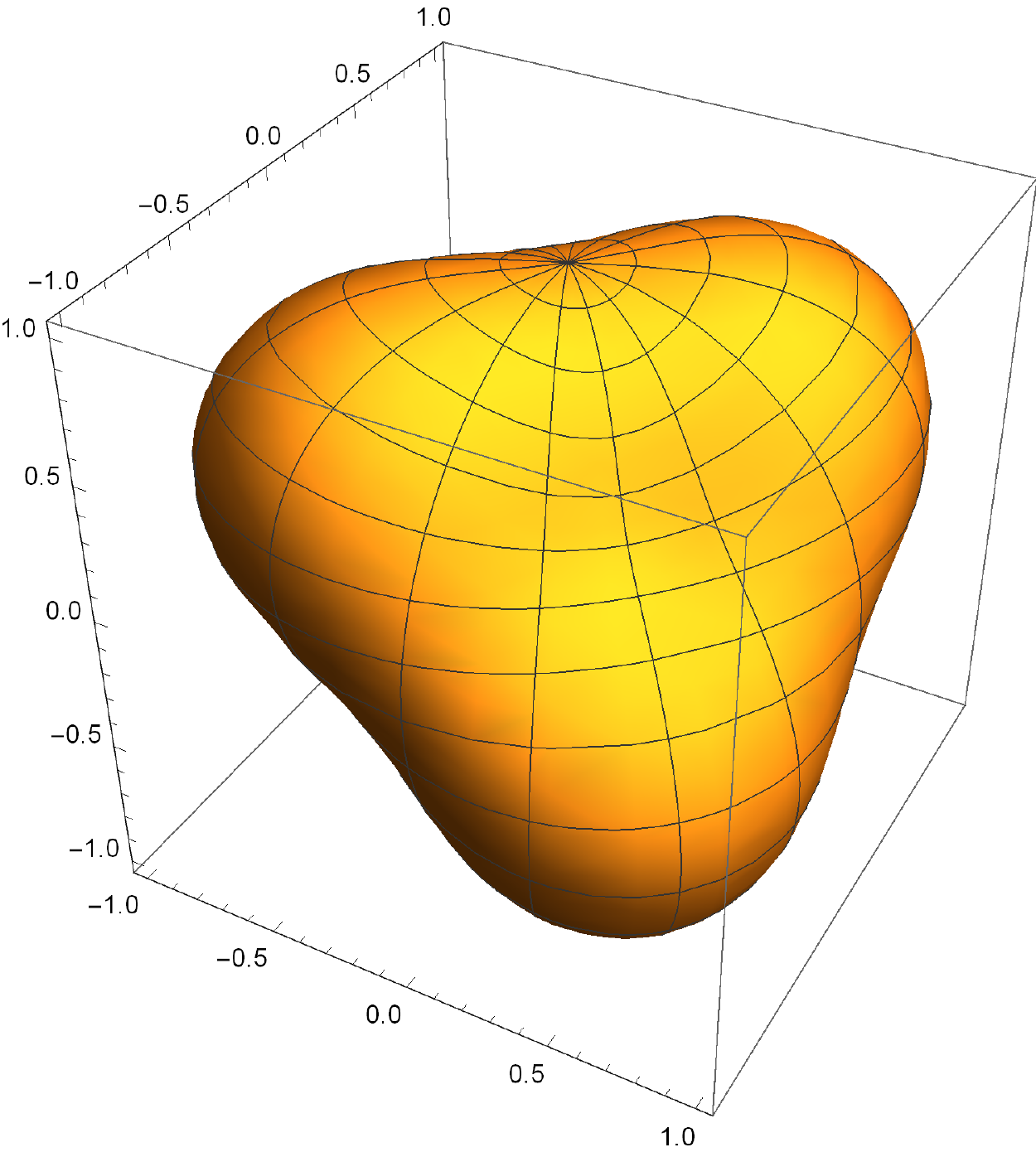}\label{fig_b}}
  \hspace*{4pt}
  \subfigure[$b= 0.5,f_z = 0.5, f_y = 0.4, f_x =0.3$]
  {\includegraphics[width=2in]{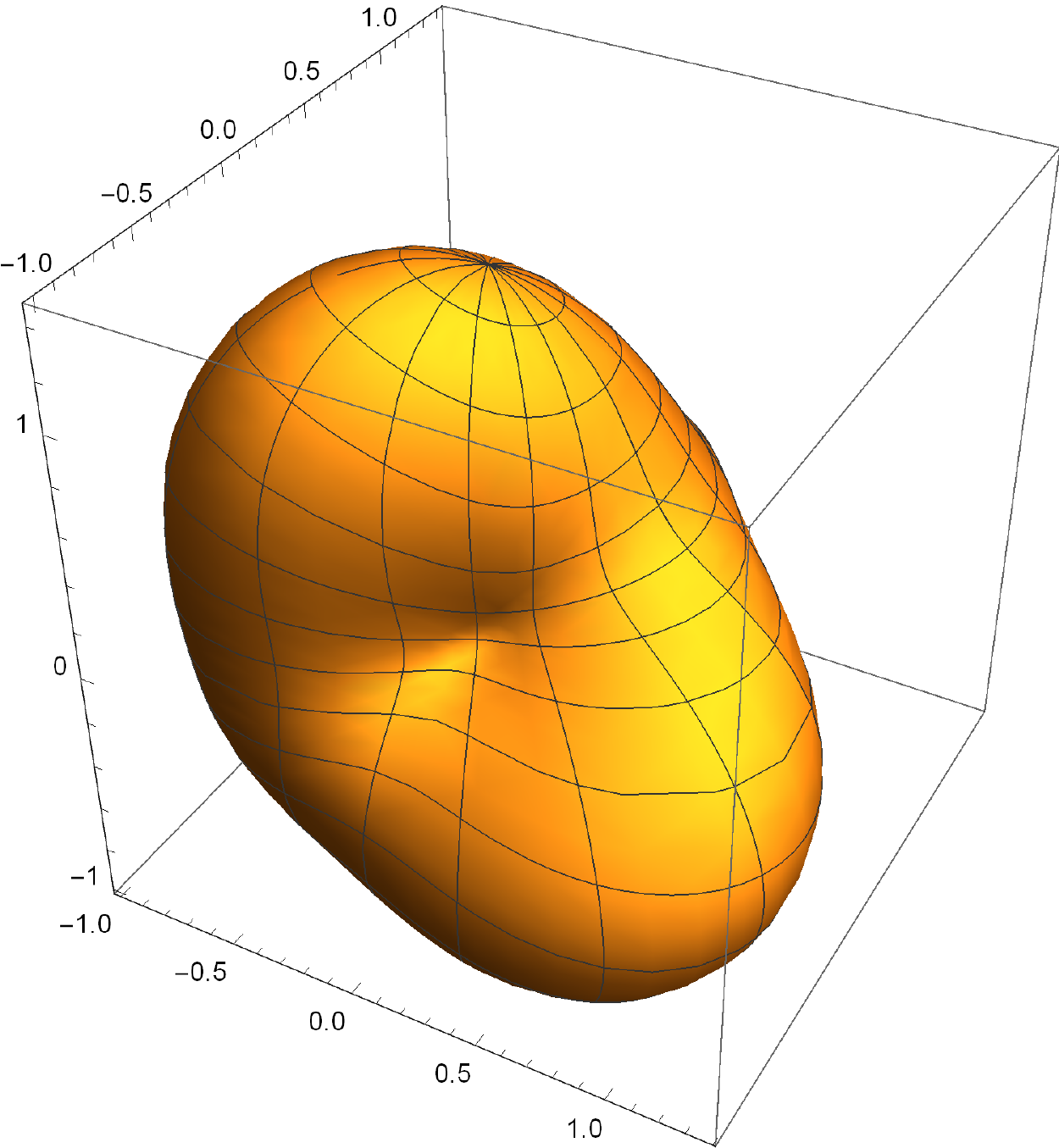}\label{fig_bfxfyfz}}}
  \caption{Examples of octupoloids with asymmetric shapes.}\label{fig_bfxyz}
\end{figure}

To learn the geometrical meaning of the deformation parameters, $b$ and $f_s$, one can investigate the shapes of the ``octupoloid" 
given by the following equation
in the spherical coordinates $R$, $\theta$, $\phi$:
\beq\label{octf}
 R(\theta ,\phi )= R_0\left[1 -A_{0}(\phi ,\theta , 0)b + \sum_{s=x,y,z}(1-2\delta _{sx})F_{0 s}(\phi ,\theta , 0 )f_{s}\right]
\eeq
in accordance with the approach of Ref.~\cite{Ham91}, bearing in mind that 
$\mathcal{D}^\lambda _{0\mu }(\phi ,\theta ,0 )=(-1)^\mu \sqrt{4\pi /(2\lambda +1)}Y_{\lambda \mu }(\theta ,\phi )$.
The shape of the octupoloid is defined by the set of deformations $b$, $f_x$, $f_y$ and $f_z$ up to the cubic group of transformations i.e. the Bohr rotations and mirror reflections 
(see Appendix \ref{OH}).
\begin{figure}[ht]
\centerline{
  \subfigure[$f_z = 0.5, b= f_x =f_y =0$]
  {\includegraphics[width=2in]{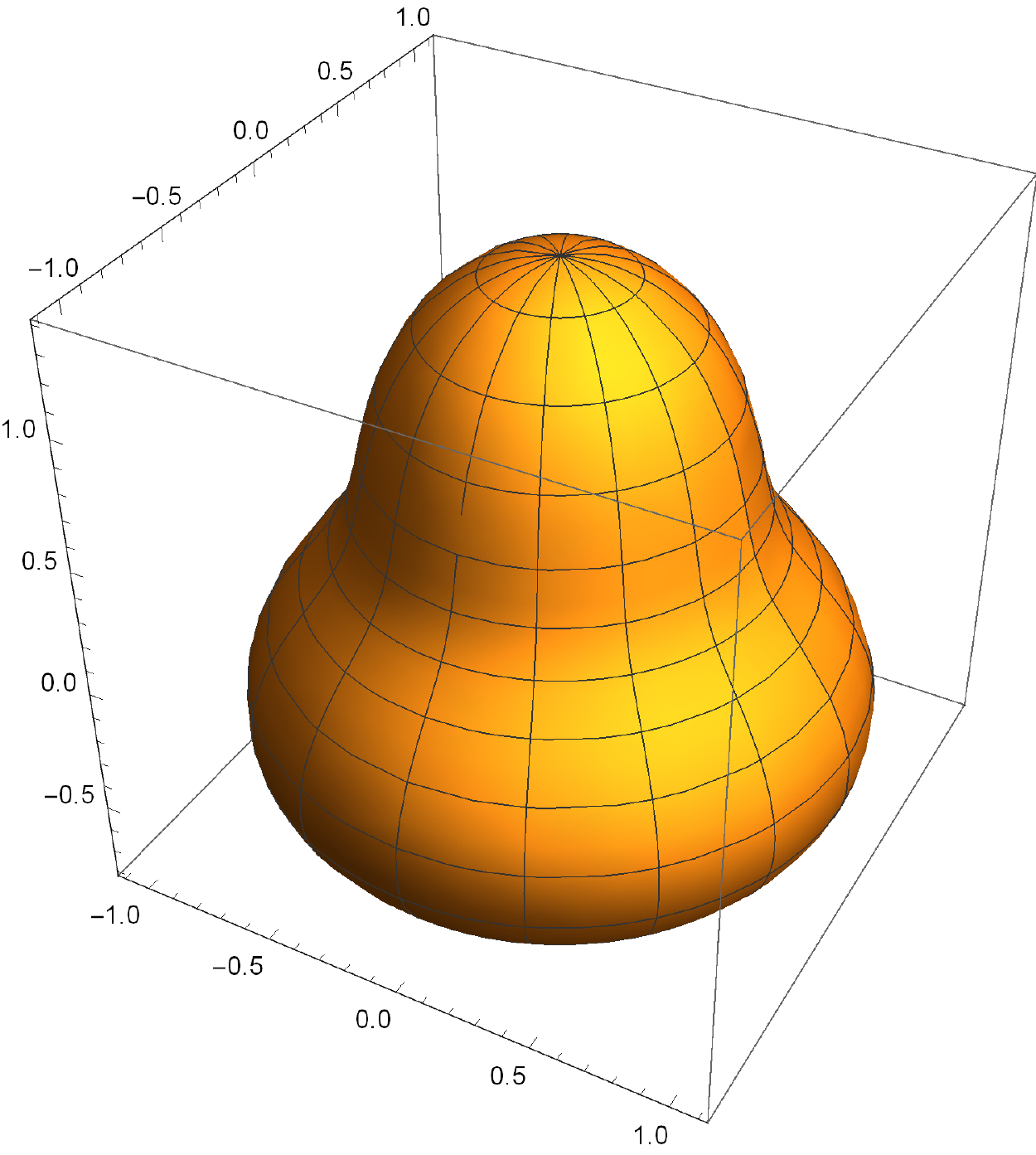}\label{fig_fz}}
  \hspace*{4pt}
  \subfigure[$f_y = 0.5, b=f_x=f_z =0$]
  {\includegraphics[width=2in]{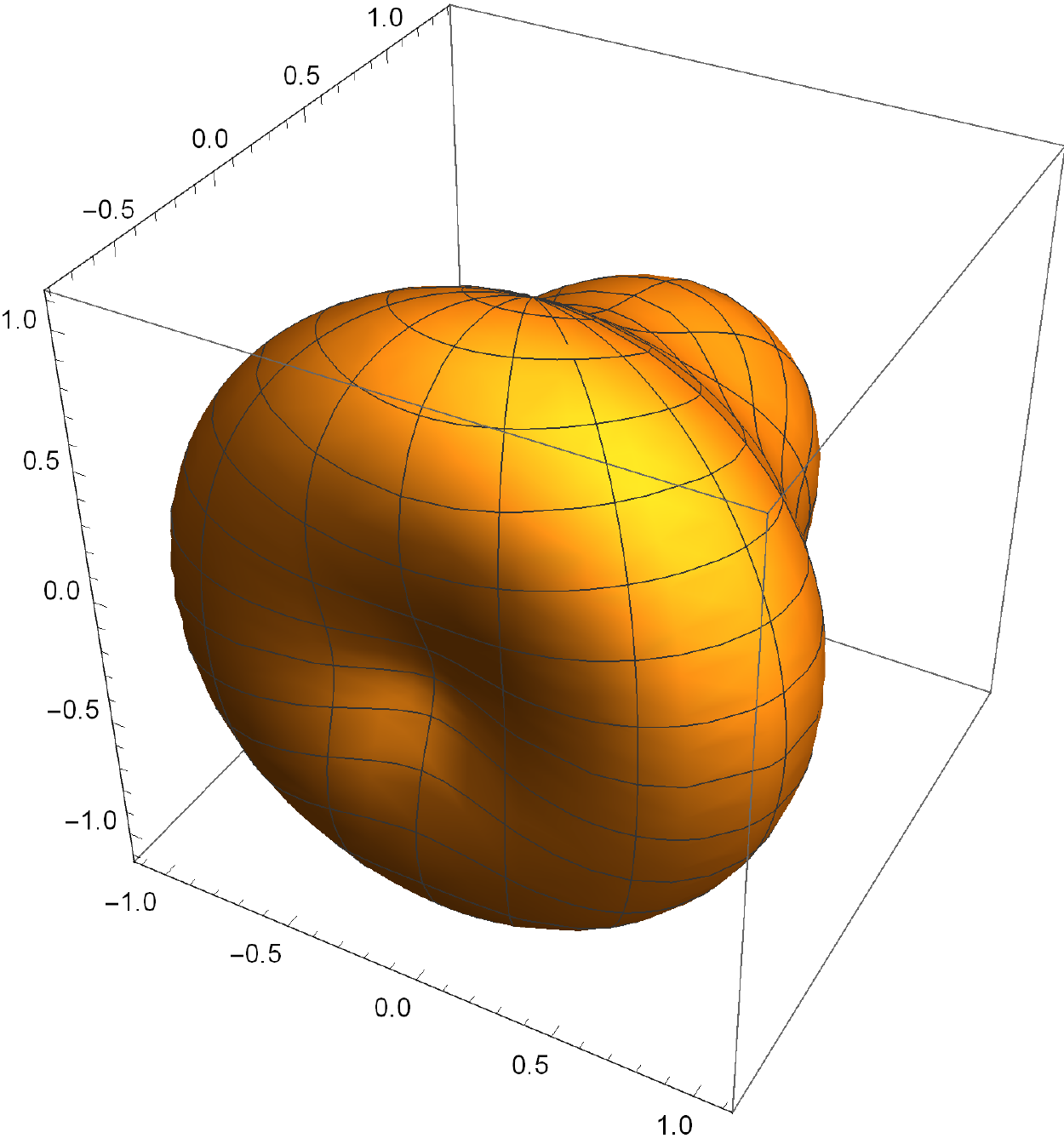}\label{fig_fy}}}
  \caption{Two octupoloids with identical axial-symmetric (pear) shapes oriented differently.}\label{fig_fyz}
\end{figure}
Two octupoloids with identical axial-symmetric (pear) shapes but oriented differently are shown in \fref{fig_fyz}. Obviously, the two sets of vector deformations then lie inside 
different pyramids of \fref{fig_pyr}. Changing the sign of the deformation gives inverted octupoloids with the same shape as that in the figure.
Examples of octupoloids with asymmetric shapes are shown in \fref{fig_bfxyz}. Their orientations and/or handedness can be changed by cubic group transormations of the deformations.
For instance, changing the signs of all the deformations gives inverted octupoloids.  
 However, when deformations belonging to one out of the two irreducible representations of the cubic group are transformed the shape of the octupoloid is changed, as shown in \fref{fig_pmbfz}.
Changing the sign of $b$ without changing the vector deformations gives a change of shape. 
 \begin{figure}[ht]
\centerline{
  \subfigure[$f_z = 0.5, b=0.5, f_x =f_y =0$]
  {\includegraphics[width=2in]{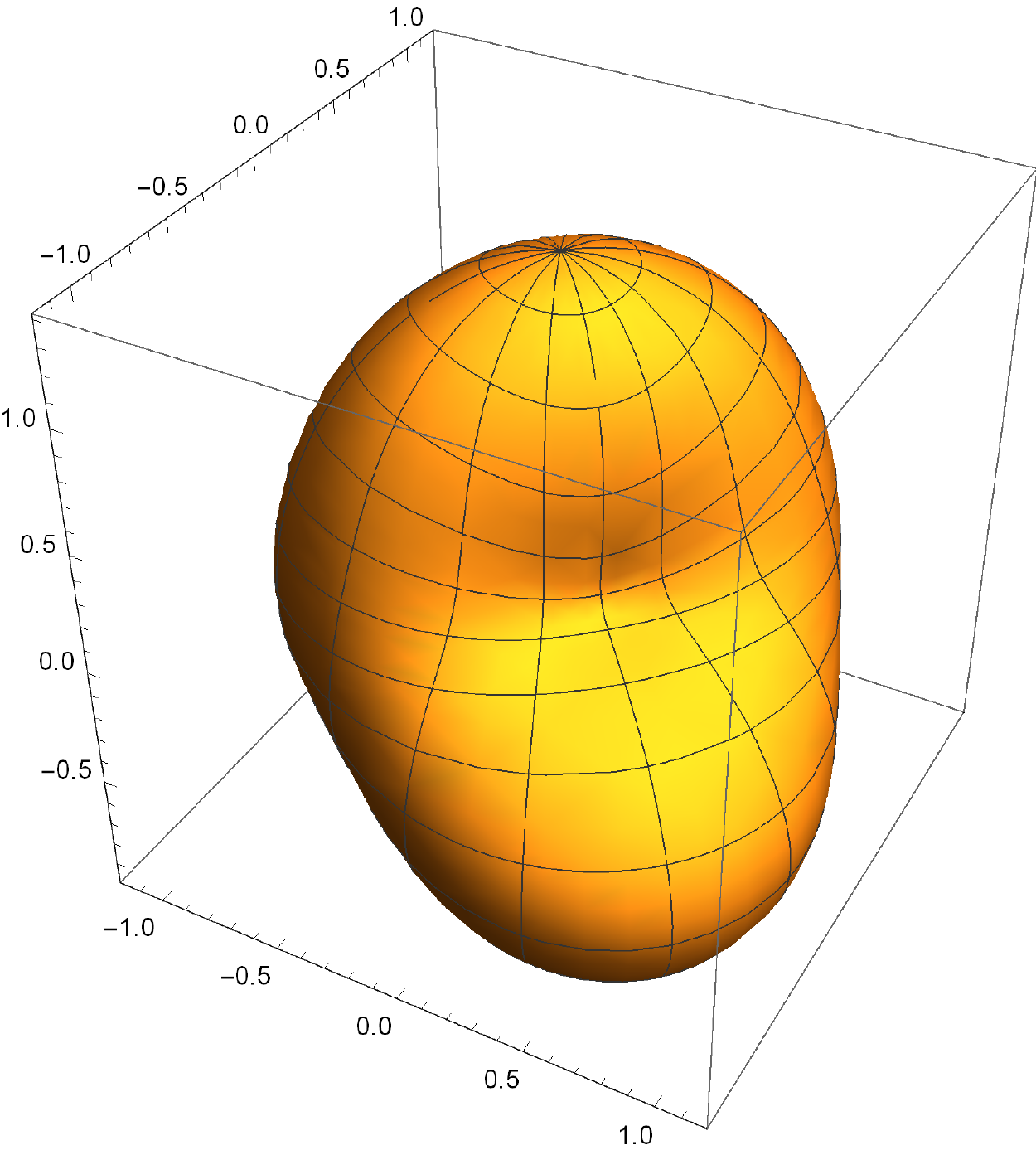}\label{fig_bfz}}
  \hspace*{4pt}
  \subfigure[$f_z = 0.5, b= -0.5, f_x=f_y =0$]
  {\includegraphics[width=2in]{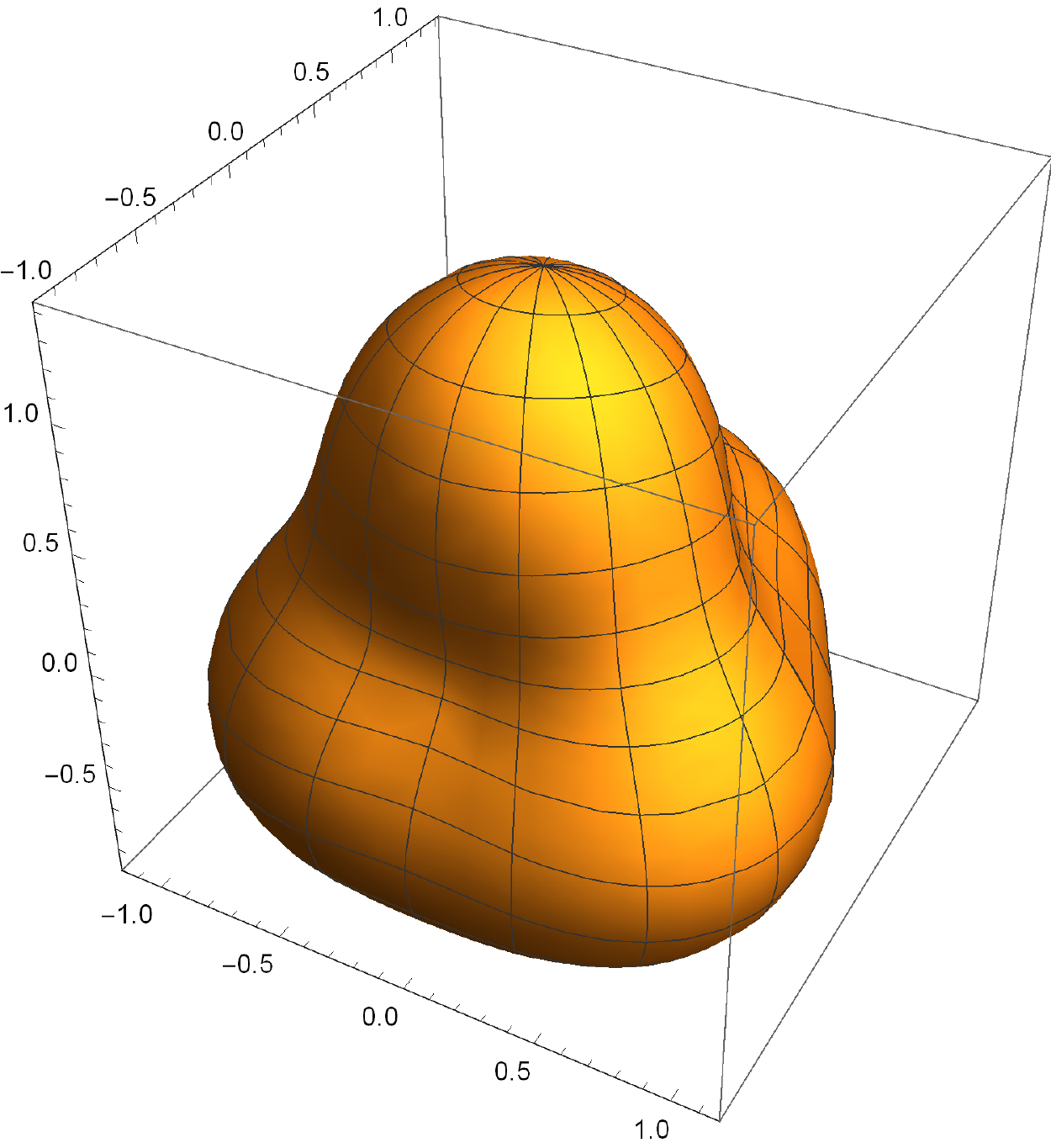}\label{fig_mbfz}}}
  \caption{Two octupoloids with opposite signs of deformation $b$  having different shapes.}\label{fig_pmbfz}
\end{figure}

\subsection{The Hamiltonian in intrinsic coordinates}\label{IHAM}

A reversible transformation between the laboratory and intrinsic coordinates
allows us to use interchangeably one or another set of variables.  The use
of intrinsic coordinates is usually 
more convenient because it gives the possibility to separate variables,
especially the Euler angles.  It is evident for potentials depending on the
laboratory coordinates through  
the elementary scalars, the number of which coincides with the number of deformations. 
Hence, the potential in \eref{haml} is a function of the number of
deformation parameters characteristic for a given multipolarity: 
two for $\lambda =2$ and four for $\lambda =3$.

To express the kinetic part of the Hamiltonian and angular momenta in intrinsic coordinates we have to convert derivatives with respect to the laboratory coordinates into derivatives with respect to the intrinsic variables.
As might be expected (cf. Ref.~\cite{Edm57}),
independently of the multipolarity of the collective space
the components of the angular momentum of \eref{angmom} can be expressed as
\beq\label{transmom}
L^{(\lambda )}_{1\kappa}= D^{1(+)}_{\kappa 0}(\omega )L_z (\omega  ) 
 -{D}^{1(+)}_{\kappa 1}(\omega ) L_x (\omega  ) 
  + {D}^{1(-)}_{\kappa 1}(\omega) L_y (\omega  )
\eeq
where $L_x$, $L_y$, $L_z$ are the Cartesian components of the intrinsic angular momentum 
depending on the Euler angles and their derivatives
only and are given by the following standard formulae: (cf. e.g. Eq.~(2.15) in Ref.~\cite{Pro09}).
\bn
{L}_x(\varphi ,\vartheta ,\psi ) &=&-i\left(-\frac{\cos{\psi}}{\sin{\vartheta}}\frac{\partial}{\partial\varphi}
+\sin{\psi}\frac{\partial}{\partial\vartheta}+\cot{\vartheta}\cos{\psi}\frac{\partial}{\partial\psi}\right) \nonumber \\
{L}_y(\varphi ,\vartheta ,\psi ) &=&-i\left(\frac{\sin{\psi}}{\sin{\vartheta}}\frac{\partial}{\partial\varphi}
+\cos{\psi}\frac{\partial}{\partial\vartheta}-\cot{\vartheta}\sin{\psi}\frac{\partial}{\partial\psi}\right) \label{inangmom} \\
{L}_z(\varphi ,\vartheta ,\psi ) &=&-i\frac{\partial}{\partial\psi} \nonumber
\en

 For $\lambda =2$ the procedure for converting the derivatives is well known and is presented in detail, e.g., in Ref.~\cite{Pro09}. A general quadrupole Hamiltonian expressed in 
intrinsic variables is divided into
the vibrational part depending only on two deformations and the rotational Hamiltonian which contains the angular momenta  $L_x$, $L_y$, $L_z$ and the deformation dependent moments of inertia.
 The intrinsic axes are always the principal axes of the tensor of inertia. When we take the kinetic energy with the Bohr inversed inertial bitensor (\ref{bibt}) the structure of the Hamiltonian
 will not change much. Namely, the mixed term in the vibrational kinetic energy will vanish and the remaining five kinetic-energy  terms ---  vibrational and rotational --- contain one 
common constant mass parameter $B_2$ instead of different deformation-dependent inertial functions. 

The procedure for transforming the octupole collective Hamiltonian to intrinsic coordinates is more involved than that for the case of $\lambda =2$. The first step of the procedure is 
conversion of the derivatives with respect
to the laboratory coordinates into those with respect to the intrinsic variables. To do this one should calculate the derivatives of the intrinsic with respect to the laboratory coordinates.
The transformation the reverse of that of \eref{intrtolabf} can be presented in the following entangled form:
\bn\label{revtransf}
&& b=\sum_\mu (A_\mu(\omega ))^*\alpha _{3\mu}, \nonumber \\
&& f_{s} = \sum_\mu (F_{\mu s}(\omega ))^*\alpha _{3\mu} \nonumber \\
&& 0= \sum_\mu (G_{\mu s}(\omega ))^*\alpha _{3\mu}
\en
for $s= x,y,z$.
Derivatives of deformations $b$ and $f_{s}$ with respect to the laboratory variables $\alpha _{3\mu}$ are equal to
\bn\label{dfdal}
&& \frac{\partial b}{\partial \alpha _{3\mu}} = (A_\mu(\omega ))^* +\sum_{\nu ,\omega}\frac{\partial}{\partial\omega}(A_\nu(\omega ))^* \alpha _{3\nu}
\frac{\partial \omega}{\partial \alpha _{3\mu}}                                       \nonumber \\
&& \frac{\partial f_{s}}{\partial \alpha _{3\mu}} = (F_{\mu s}(\omega ))^* + \sum_{\nu ,\omega}\frac{\partial}{\partial\omega}(F_{\nu s}(\omega ))^* \alpha _{3\nu}
\frac{\partial \omega}{\partial \alpha _{3\mu}}                                      
\en
Derivatives of the Euler angles with respect to $\alpha _{3\mu}$ are obtained by solving the following set of linear equations:
\bn\label{eqforder}
&& \sum_\nu \left[(\frac{\partial}{\partial \varphi}(G_{\nu s}(\omega ))^*)\frac{\partial \varphi}{\partial \alpha _{3\mu}}+
(\frac{\partial}{\partial \vartheta}(G_{\nu s}(\omega ))^*)\frac{\partial \vartheta}{\partial \alpha _{3\mu}}+
(\frac{\partial}{\partial \psi}(G_{\nu s}(\omega ))^*)\frac{\partial \psi}{\partial \alpha _{3\mu}}\right]\alpha _{3\nu}  \nonumber \\
&& +(G_{\mu s}(\omega ))^*=0 \qquad \mathrm{for} \quad s=x,y,z.
\en
To calculate all the derivatives from Eqs.~(\ref{dfdal}) and(\ref{eqforder}) one should calculate the derivatives of the octupole cubic Wigner functions with respect to the Euler angles.
Using Eqs.~(\ref{semicart}) and (\ref{cubic}) the derivatives of the cubic functions can be expressed by derivatives of the Wigner functions themselves. E.g., handbook~\cite{Var88} 
contains formulae for the derivatives in question and other relevant properties of the Wigner functions\footnote{Note that the Wigner functions from Ref.~\cite{Var88} are the complex 
conjugate of those used here.}.
Finally, solutions of \eref{eqforder} for derivatives of the Euler angles with respect to $\alpha _{3\mu}$ are:
\bn
&& \frac{\partial\varphi}{\partial\alpha _{3\mu}}= \frac{1}{2d_f(b,f_s)}\frac{1}{\sin{\vartheta}}\left[-\cos{\psi}(\Gamma_{\mu x}(b,f_s,\omega ))^*+\sin{\psi}(\Gamma_y(b,f_s ,\omega ))^*\right]
     \nonumber \\
&& \frac{\partial\vartheta}{\partial\alpha _{3\mu}}= \frac{1}{2d_f(b,f_s)}\left[\sin{\psi}(\Gamma_{\mu x}(b,f_s ,\omega ))^*+\cos{\psi}(\Gamma_y(b,f_s ,\omega ))^*\right] \nonumber \\
&&  \frac{\partial\psi}{\partial\alpha _{3\mu}}= \frac{1}{2d_f(b,f_s)}(\Gamma_{\mu z}(b,f_s ,\omega ))^* -\cos{\vartheta}\frac{\partial\varphi}{\partial\alpha _{3\mu}} \label{domdal}
\en
where
\bn\label{gam}
&& \Gamma _{\mu s}(b,f_x,f_y,f_z,\omega ))^* = \frac{15}{16}
    f_{s}\left((G_{\mu s}(\omega ))^* f_{s}-(G_{\mu t}(\omega ))^* f_{t}-(G_{\mu u}(\omega ))^* f_{u}\right) \nonumber \\
&&    -b^2(G_{\mu s}(\omega ))^* +\frac{\sqrt{15}}{4}b\left(f_{u} (G_{\mu t}(\omega ))^*+f_{t}(G_{\mu u}(\omega ))^*\right)  \omega ,
\en
and $s,t,u$ are circular permutations of $x,y,z$ here and further below.
Using Eqs.~(\ref{inangmom}), (\ref{dfdal}) and (\ref{domdal}) we are in a position to
express derivatives with respect to the laboratory  by derivatives
with respect to the intrinsic coordinates in the following two equivalent ways:
\bn \label{dlab}
&& \frac{\partial}{\partial\alpha _{3\mu}}= (A_\mu(\omega ))^*\frac{\partial}{\partial b} + \sum_{s=x,y,z}\left\{(F_{\mu s}(\omega ))^* \frac{\partial}{\partial f_{s}}\right.
  \nonumber \\
&& + \left.\frac{i}{2d_f(b,f_x,f_y,f_z)}{(\Gamma _{\mu s}(b,f_x.f_y,f_z ,\omega ))^*
\left[L_s (\omega )- J^{(f)}_s (f_t,f_u )
 \right]}
\right\} \nonumber \\
&& =\left[\frac{\partial}{\partial b} + \frac{1}{d_f(b,f_x,f_y,f_z)}\frac{\partial d_f(b,f_x,f_y,f_z)}{\partial b} 
\right](A_\mu(\omega ))^* \nonumber \\
&& \sum_{s=x,y,z}\left\{ \left[\frac{\partial}{\partial f_{s}} + \frac{1}{d_f(b,f_x,f_y,f_z)}\frac{\partial d_f(b,f_x,f_y,f_z)}{\partial f_s} \right)\right](F_{\mu s}(\omega ))^* \nonumber \\
&& + \left.\frac{i}{2d_f(b,f_x,f_y,f_z)}
\left[L_s (\omega )- J^{(f)}_s (f_t ,f_u )\right]
 (\Gamma _{\mu s}(b,f_x,f_y,f_z ,\omega ))^* 
 \right\}
\en
where the differential operators 
\beq\label{vibj}
J^{(f)}_s (f_t, f_u )= \frac{3}{2}i\left(f_{t}\frac{\partial}{\partial f_{u}} - f_{u}\frac{\partial}{\partial f_{t}}\right) 
\eeq
stand for angular momenta carried by vibrations of the octupole vector deformations $f_x,\ f_y,\ f_z$ and will be called the octupole vibrational angular momenta.

Using both versions of the right-hand side of \eref{dlab} and taking advantage of the unitarity of the octupole cubic Wigner functions we are able to express the Hamiltonian of 
\eref{haml} for $\lambda =3$
in the intrinsic coordinates. To observe the inherent characteristics of the octupole collective motion an octupole Hamiltonian with the simplest inverse inertial bitensor, namely that of \eref{bibt},
will be presented here. This Hamiltonian when expressed in the Euler angles and octupole deformations is as follows:
\bn\label{intrham}
&& H_3(b, f_x,f_y,f_z , \omega)= -\frac{1}{2 B_3 d_f (b,f_x,f_y,f_z )}\left\{\frac{\partial}{\partial b}d_f (b,f_x,f_y,f_z )\frac{\partial}{\partial b}\right. \nonumber \\
&&          +\left.\sum_{s}\frac{\partial}{\partial f_s}d_f (b,f_x,f_y,f_z )\frac{\partial}{\partial f_s} 
             -\sum_{s,s'}(L_s (\omega )-J^{(f)}_s (f_t,f_u ))\right.\nonumber \\
&& \left. \times d_f (b,f_x,f_y,f_z )(\hat{I}^{(f)}(b,f_x,f_y,f_z ))^{-1}_{ss'} 
  \times (L_{s'} (\omega )-J^{(f)}_{s'} (f_{t'},f_{u'} ))\right\} \nonumber \\
&& +V(b,f_x,f_y.f_z)
\en
where the Cartesian tensor of the moments of inertia is equal to
\bn\label{inertiaf}
&& \hat{I}^{(f)}(b,f_x,f_y,f_z)= \left(\ba{ccc} 4 b^2 + \frac{15}{4}(f_y^2+f_z^2) & \frac{15}{4}f_x f_y + 2\sqrt{15} b f_z & \frac{15}{4}f_x f_z+ 2\sqrt{15} b f_y \\
                        \frac{15}{4}f_x f_y + 2\sqrt{15} b f_z & 4 b^2 + \frac{15}{4}(f_x^2+f_z^2) & \frac{15}{4}f_y f_z+ 2\sqrt{15} b f_x \\
                         \frac{15}{4}f_x f_z+ 2\sqrt{15} b f_y &  \frac{15}{4}f_y f_z+ 2\sqrt{15} b f_x & 4 b^2 + \frac{15}{4}(f_x^2+f_y^2)\ea\right) ,\nonumber \\
\en
It is seen that the intrinsic axes of an octupole system are not the principal axes of the moment of inertia as one would expect. The octupole vibrations, contrary to the quadrupole ones, carry
their own angular momentum, which interacts by the Coriolis and centrifugal interactions with the total angular momentum (cf. Ref.~\cite{Boh75}). This seems to be the most striking feature of the octupole rotations. 
In conclusion, the kinetic energy part of the Hamiltonian of \eref{intrham} consists of ten terms: the four separate vibrations and the six rotational terms. In general, 
the Hamiltonian of \eref{haml} for $\lambda =3$,
when expressed in the intrinsic variables, can contain additionally six mixed vibrational terms of type $\partial /\partial f_s\dots \partial /\partial f_{s'}$ ($s\ne s'$)
and twelve vibration-rotation terms of type $(\partial /\partial f_s \dots (L_{s'}- J^{(f)}_{s'}) + \mathrm{h.c.})$.

\subsection{Axially-symmetric deformation}\label{AS}
Many even-even nuclei show a static quadrupole deformation with axial symmetry. In the small-oscillations approximation of the collective Hamiltonian a simple picture of the quadrupole excitations 
has emerged (consult e.g. Chapter 6 in Ref.~\cite{Eis87}). Two separate intrinsic vibrations appear, namely: the $\beta $-vibration of deformation $a_{20}\approx\beta $ around the
equilibrium deformation $\beta _{\mathrm{eq}}$, and the $\gamma $-vibration ($a_{22}\approx\beta \gamma $) strongly coupled to rotations around the symmetry axis. On the other hand, rotations around axes perpendicular 
to the symmetry axis are weakly coupled to the vibrations and form characteristic rotational bands built on top of every vibrational level.

How is it in the case of a static axially-symmetric octupole deformation? The equilibrium points in the four-dimensional deformation space are then supposed to be 
$b = f_x =f_y = 0,\ f_z =\pm f_\mathrm{eq}$. The equilibrium shape is shown in \fref{fig_fz}. A rough approximation of the small oscillations around the two (because of the mirror symmetry) equilibrium
points for the Hamiltonian of \eref{intrham} (with the original Bohr kinetic energy) reads as follows:
\bn\label{smosc}
&& H_3(b, f_x,f_y,f_z , \omega) \approx \sum_{s=x,y}\left[-\frac{1}{2 B_3}\frac{\partial ^2}{\partial f_s^2} +\frac{1}{2}C_s f_s^2\right] \nonumber \\
&& \phantom{H(b, f_x,f_y,f_z , \omega)} -\frac{1}{2 B_3}\frac{\partial ^2}{\partial f_z^2} +\frac{1}{2}C_z (| f_z| -f_{\mathrm{eq}})^2 \nonumber \\
&& \phantom{H(b, f_x,f_y,f_z , \omega)} +\frac{1}{2 B_3}\left[-\frac{1}{b}\frac{\partial }{\partial b}b\frac{\partial }{\partial b} 
+ \frac{(L_z (\omega )-J^{(f)}_z (f_x,f_y ))^2}{4 b^2}\right] +\frac{1}{2}C_b b^2 \nonumber \\
&& \phantom{H(b, f_x,f_y,f_z , \omega)} +\sum_{s=x,y}\frac{2}{15 B_3 f_{\mathrm{eq}}^2}(L_s (\omega )-J^{(f)}_s (f_t,f_u ))^2
\en
From the form of the Hamiltonian given above the following pattern of the small oscillations around the axial-symmetric octupole shape emerges, namely:
\begin{itemlist}
\item Two harmonic oscillations in coordinates $f_x$ and $f_y$ with stiffnesses $C_x$ and $C_y$, respectively (x- and y-vibrations),
\item The double-oscillator z-vibrations around points $f_z = \pm f_{\mathrm{eq}}$ with stiffness $C_z$ (cf. Ref.~\cite{Merz61}),
\item The b-vibration with stiffness $C_b$ strongly coupled  to rotation around the symmetry axis z,
\item Rotations around the x- and y-axes perpendicular to the symmetry axis with constant moment of inertia equal to $(15/4)B_3 f_{\mathrm{eq}}^2$.
\end{itemlist}
The rotations around the x and y axes form rotational bands on top of vibrational levels. However, the bands are disturbed by the Coriolis interaction, being a kind of
the rotation-vibration interaction. In turn, centrifugal forces affect the four separate vibrations and can mix them with each other.

\section{Conclusion}\label{CONCL}
In the previous Sections a formalism for the octupole collective Hamiltonian has been presented and compared to that for the well-known quadrupole one.
For a few reasons, like a number of degrees of freedom greater by two, negative parity, additional simplifications in the quadrupole case, the theory of the octupole collective 
Hamiltonian is essentially more complicated, and therefore less developed than that of the quadrupole collective motion. A substantial feature of the octupole motion, which does not seem 
to be realized,
is that the intrinsic vector x-, y- and z-vibrations carry a non-zero angular momentum. This is obviously not the case for the quadrupole $\beta $- and $\gamma $-vibrations.   

Obviously, a realistic collective model should take into account both modes, the quadrupole and the octupole together \cite{Roh82}. A separate consideration of 
the $\lambda =2$ and $\lambda =3$ cases either serves
as a tool for developing a formalism and methods of treatment, or is an approximation. When we take, for instance, the kinetic energies of both modes with the Bohr inverse inertial bitensors of 
\eref{bibt}, the total quadrupole-octupole Hamiltonian is the sum of the kinetic energies parametrized by the two mass parameters, $B_2$ and $B_3$, respectively, and 
the potential $V(\bls{\alpha }_2 ,\bls{\alpha }_3 )$,
which can contain a possible quadrupole-octupole interaction. However, modern collective Hamiltonians are extracted from microscopic theories, which seem to give inverse inertial bitensors
$B^{-1}_{\lambda \mu \lambda \mu '}(\bls{\alpha }_2 ,\bls{\alpha }_3 )$ dependent on both sets of coordinates for both $\lambda $'s. Then, for instance, assumption no. (6) from \sref{GCH} that
the collective Hamiltonians contain isotropic functions of coordinates, is not valid. In consequence, the bitensor $B^{-1}_{2 \mu 2 \mu '}(\bls{\alpha }_2 ,\bls{\alpha }_3 )$ can have more than six 
independent components. Furthermore, it is natural to allow for the appearance of mixed quadrupole-octupole terms in the total Hamiltonian. These terms would have the following form:
 \bn\label{mixham}
 H_{23}(\bls{\alpha }_2,\bls{\alpha }_3 )&=& -\frac{1}{2 W(\bls{\alpha }_2,\bls{\alpha }_3) }\left[\sum_{\mu ,\nu}\frac{\partial}{\partial\alpha _{2 \mu}}
W(\bls{\alpha }_2,\bls{\alpha }_3 ) B^{-1}_{2 \mu 3 \nu}(\bls{\alpha }_2,\bls{\alpha }_3 )\frac{\partial}{\partial\alpha _{3 \nu}}\right. \nonumber \\ 
 &+& \left.\sum_{\mu ,\nu}\frac{\partial}{\partial\alpha _{3 \mu}}
W(\bls{\alpha }_2,\bls{\alpha }_3 ) B^{-1}_{3 \mu 2 \nu}(\bls{\alpha }_2,\bls{\alpha }_3 )\frac{\partial}{\partial\alpha _{2 \nu}}\right]
+ V_{23}(\bls{\alpha }_2,\bls{\alpha }_3  ) .
\en
Should $H_{23}$  be invariant under space inversion and Hermitian,  $B^{-1}_{\lambda  \mu \lambda ' \mu '}(\bls{\alpha }_2,\bls{\alpha }_3 )$ is symmetric 
($ \lambda  \mu \rightleftharpoons \lambda ' \mu '$) and odd.
By analogy to \eref{bll} the mixed bitensor can be presented in the following form:
\beq\label{b23}
B^{-1}_{\lambda \mu \lambda ' \mu '}(\bls{\alpha }_2 ,\bls{\alpha }_3  ) = \sum_{l  =0}^{2 }(\lambda \mu \lambda ' \mu '|2l + 1  m )T_{2l +1  m }(\bls{\alpha }_2 ,\bls{\alpha }_3 )
\eeq
for $\lambda \ne\lambda ' = 2 ,3$. Tensors $T_{2l + 1 m}$ should have negative parity. The bitensor of \eref{b23} has 21 components altogether.

When the quadrupole-octupole collective Hamiltonian is considered assumptions nos. (1) -- (6) from \sref{GCH} should be extended to the collective space of both tensors,
$\bls{\alpha }_2$ and $\bls{\alpha }_3$. The practical role of assumption no. (6) is that no material tensors appear for the nuclear medium.   
Under these extended assumptions the most general form of a quadrupole-octupole Hamiltonian reads as follows: 
 \bn\label{totham}
 H(\bls{\alpha }_2,\bls{\alpha }_3 )&=& -\frac{1}{2 W(\bls{\alpha }_2,\bls{\alpha }_3) }\left[\sum_{\lambda ,\lambda ' =2,3}\sum_{\mu , \mu '}\frac{\partial}{\partial\alpha _{\lambda  \mu}}
W(\bls{\alpha }_2,\bls{\alpha }_3 ) B^{-1}_{\lambda  \mu \lambda ' \mu '}(\bls{\alpha }_2,\bls{\alpha }_3 )\frac{\partial}{\partial\alpha _{\lambda ' \mu '}}\right] \nonumber \\ 
 &+& V(\bls{\alpha }_2,\bls{\alpha }_3  ) .
\en
It is parametrized by 64 coordinate-dependent inertial functions being components of the three inverse inertial bitensors, scalar weight and potential. The weight can possibly be equal to the square root of the determinant of the $12\times 12$ matrix of components of the inertial bitensors. The potential is a function of the coordinates through nine scalars described as deformations. 
In order to separate these nine variables from the twelve coordinates,
a body-fixed intrinsic frame of reference
and intrinsic coordinates have to be introduced. One can do this in different ways. For instance, the principal axes of tensor $\bls{\alpha }_2$ oriented by the three Euler angles with respect to the laboratory axes can be treated as the intrinsic axes (cf. Ref.~\cite{Roh82}). Then the two remaining intrinsic components of $\bls{\alpha }_2$ and all
the seven intrinsic components of $\bls{\alpha }_3$ can be considered as deformations. Another way is to exchange the roles of tensors $\bls{\alpha }_2$ and 
$\bls{\alpha }_3$ and consider one of the frames defined in \sref{IF} through the octupole tensor as the intrinsic frame. One can also define two intrinsic frames for tensors 
$\bls{\alpha }_2$ and $\bls{\alpha }_3$ separately and treat the rotation of one frame with respect to the other as an intrinsic motion. Then the relative Euler angles have
the status of deformations. Finally, in the case of a weak and well separated interaction between the modes one can treat them separately and then diagonalize the interaction within the product basis. 

Only recently an attempt to solve a quadrupole-octupole model, similar to that presented here, however not based on Hamiltonian (\ref{totham})
and with a restricted number of degrees of freedom, has been undertaken \cite{Dob16}. The model has been applied to the positive- and negative-parity collective levels of the $^{156}$Gd nucleus. 
In any case, the problem of the full quadrupole-octupole collective Hamiltonian is apparently complicated enough and still awaits practical applications to the spectroscopy
of nuclear collective excitations.

\section*{Acknowledgments}
The subject of the present paper was initiated some time ago by the first author (S.G.R.), then a Fellow of the Alexander von Humboldt Foundation, in collaboration with Walter Greiner, the then Director of the 
{\it Institut f\"ur Theoretische Physik der Johann Wolfgang Goethe-Universit\"at}. \\
The second author (L.P.) gratefully acknowledges support from the Polish National Science Center (NCN), grant no. 2013/10/M/ST2/00427.

\begin{appendix}[Isotropic tensor fields in the octupole collective space]\label{OTF}
\section{The positive spin-parity elementary tensors}\label{ET}
The twelve positive spin-parity elementary tensors of \eref{elten} in the octupole collective space are as follows:
\bn\label{eet}
 l = 0 && \quad \bls{t}_0^{(2)},\; \bls{t}_0^{(4)},\; \bls{t}_0^{(6)},\; \bls{t}_0^{(10)}, \nonumber \\ 
 l = 1 && \quad \bls{t}_1^{(3)},\; \bls{t}_1^{(5)},\; \bls{t}_1^{(7)}, \nonumber  \\
 l = 2 && \quad \bls{t}_2^{(2)}, \; \bls{t}_2^{(4)}, \nonumber \\
 l = 3 && \quad \bls{t}_3^{(1)}, \; \bls{t}_3^{(3)}, \nonumber \\
 l = 4 && \quad \bls{t}_4^{(2)}.
\en

\section{Independent fundamental even tensors}\label{IFT}
Sets of fundamental even tensors with even ranks from 0 to 6 are listed in \tref{fet6}. The choice of independent tensors need not be unique. This is because
all fundamental tensors of a given rank (all possible alignments of the elementary tensors) are related to each other through a number of syzygies which can eliminate this or that tensor.
\begin{table}[h]
\tbl{Fundamental even tensors with ranks $\Lambda = 0,\ 2,\ 4,\ 6 $}
{\btab{@{}lll@{}}\toprule
$\Lambda $& $k_{3\Lambda }$ & $\bls{\tau }_{k\Lambda }(\bls{\alpha }_3)$ for $k = 1,\dots ,k_{3\Lambda }$ \\
\colrule
0 & 1 & 1 \\
2 & 5 & $\bls{t}_2^{(2)}, \;\bls{t}_2^{(4)}, \; [\bls{t}_1^{(3)}\times \bls{t }_1^{(3)}]_2 ,\;
  [\bls{t}_1^{(3)}\times \bls{t }_1^{(5)}]_2 ,\; [\bls{t}_1^{(3)}\times \bls{t}_1^{(7)}]_2 $\\
4 & 9 & $\bls{t}_4^{(2)},\; [\bls{t}_3^{(1)}\times \bls{t}_1^{(3)}]_4,\; [\bls{t}_2^{(2)}\times \bls{t}_2^{(2)}]_4 ,
 \; [\bls{t}_2^{(2)}\times \bls{t}_2^{(4)}]_4,\; [\bls{t}_3^{(1)}\times \bls{t}_1^{(5)}]_4 ,$ \\
 & & $[\bls{t}_2^{(2)}\times \bls{t}_1^{(3)}\times \bls{t}_1^{(3)}]_4,\; [\bls{t}_3^{(1)}\times \bls{t}_1^{(7)}]_4 ,\; 
[\bls{t}_2^{(2)}\times \bls{t}_1^{(3)}\times \bls{t}_1^{(5)}]_4,\; [\bls{t}_2^{(2)}\times\bls{t}_3^{(1)}\times \bls{t}_1^{(7)}]_4 $\\
6 & 13 & $[\bls{t}^{(1)}_3\times\bls{t}^{(1)}_3]_6 ,\; [\bls{t}^{(2)}_2\times\bls{t}^{(2)}_4]_6 ,\; 
 [\bls{t}^{(3)}_3\times\bls{t}^{(1)}_3]_6 , \;   [\bls{t}^{(4)}_2\times\bls{t}^{(2)}_4]_6 ,\;
[\bls{t}^{(3)}_1\times\bls{t}^{(2)}_2\times\bls{t}^{(1)}_3]_6$ \\
& &  $[\bls{t}^{(3)}_1\times\bls{t}^{(3)}_1\times\bls{t}^{(2)}_4]_6 ,\;  [\bls{t}^{(5)}_1\times\bls{t}^{(2)}_2\times\bls{t}^{(1)}_3]_6 ,\;
[\bls{t}^{(4)}_2\times\bls{t}^{(2)}_2\times\bls{t}^{(2)}_2]_6 ,\; 
    [\bls{t}^{(3)}_1\times\bls{t}^{(5)}_1\times\bls{t}^{(2)}_4]_6 $ \\
& &  $[\bls{t}^{(7)}_1\times\bls{t}^{(2)}_2\times\bls{t}^{(1)}_3]_6 ,\; [\bls{t}^{(3)}_1\times\bls{t}^{(3)}_1\times\bls{t}^{(3)}_1\times\bls{t}^{(1)}_3]_6 ,\; 
 [\bls{t}^{(4)}_2\times\bls{t}^{(7)}_1\times\bls{t}^{(1)}_3]_6 ,\; [\bls{t}^{(3)}_1\times\bls{t}^{(7)}_1\times\bls{t}^{(2)}_4]_6 $\\
 \botrule
 \etab}\label{fet6}
 \end{table}
 
\end{appendix}
\begin{appendix}[Symmetries of the coordinate frame]\label{SCF}
\section{Cubic holohedral group}\label{OH}
The cubic holohedral  O$_\mathrm{h}$ group is a natural symmetry group of the
three-dimensional coordinate system because the forty-eight group elements
are: the eight reverses of the axis arrows
for each out of six permutations of axes.
The three Bohr rotations, $R_1,\ R_2,\ R_3$, and the inversion $P$ can serve as
generators of this group (see Ref.~\cite{Pro09} and Sect.~4.4 in Ref.~\cite{Eis87}).
In general, the O$_\mathrm{h}$ group has ten irreducible representations \cite{Ham64}, namely
\begin{itemlist}
\item four one-dimensional, denoted as A$^\pm_1$, A$^\pm_2$,
\item two two-dimensional, denoted as E$^\pm$,
\item four three-dimensional, denoted as F$^\pm_1$, F$^\pm_1$.
\end{itemlist}
The tensor representations D$^\lambda $ of the O(3) orthogonal group can be
 decomposed into the following irreducible representations of O$_\mathrm{h}$, namely
 \begin{itemlist}
 \item  irreps E$^+$, F$^+_2$ for $\lambda =2$,
 \item  irreps A$^-_2$, F$^-_1$, F$^-_2$ for $\lambda =3$. 
 \end{itemlist}
 \section{Cubic coordinates}\label{CC}
The decomposition of the real and imaginary parts, $a_{\lambda \mu },\ b_{\lambda \mu }$ of the spherical components of tensors $\bls{\alpha }_\lambda $ into cubic coordinates is:
\begin{itemlist}
\item $\lambda =2$
\bn\label{cc2}
\mathrm{E}^+\left\{ \ba{c} e_{20} =a_{20}\\ e_{22} = a_{22}, \ea\right. &\quad & \mathrm{F}^+_2\left\{ \ba{c} g_{2x} = - b_{21}, \\ g_{2y} = - a_{21}, \\ g_{2z}= b_{22}, \ea \right.
\en 
\item $\lambda = 3$
\bn\label{cc3}
&& \mathrm{A}^-_2 : \; b_3\equiv b = b_{32} ,  \nonumber  \\
&& \mathrm{F}^-_1\left\{ \ba{c} f_{3x}\equiv f_{x}= \sqrt{\frac{3}{8}}a_{31}-\sqrt{\frac{5}{8}}a_{33}, \\
 f_{3y}\equiv f_{y}= \sqrt{\frac{3}{8}}b_{31}+\sqrt{\frac{5}{8}}b_{33}, \\
f_{3z} \equiv  f_{z}= a_{30}, \ea\right. \nonumber \\
&& \mathrm{F}^-_2 \left\{ \ba{c} g_{3x}\equiv g_{x}= \sqrt{\frac{5}{8}}a_{31}+\sqrt{\frac{3}{8}}a_{33}, \\
 g_{3y}\equiv g_{y}= -\sqrt{\frac{5}{8}}b_{31}+\sqrt{\frac{3}{8}}b_{33}, \\
 g_{3z}\equiv g_{z}= a_{32}.\ea\right.
 \en 
\end{itemlist}            
Curly brackets match the cubic coordinates belonging to given irreducible representations of O$_\mathrm{h}$.
\end{appendix}

\end{document}